\PassOptionsToPackage{hyphens}{url}
\PassOptionsToPackage{shortlabels,inline}{enumitem}
\documentclass[10pt,twocolumn,letterpaper]{article}

\usepackage[utf8]{inputenc}
\usepackage[font=small]{caption}
\usepackage{amsmath,amssymb,amsthm,nccmath,mathtools} 
\usepackage{booktabs,multirow,adjustbox}             
\usepackage{algorithm,algpseudocode}                 
\usepackage[switch]{lineno}
\usepackage{microtype}
\usepackage{graphicx}
\usepackage{academicons}
\usepackage{csquotes}
\usepackage[dvipsnames]{xcolor} 
\usepackage{lipsum}
\usepackage{subcaption}
\usepackage{dblfloatfix}
\usepackage{enumitem} 
\usepackage{cvpr}              
\definecolor{cvprblue}{rgb}{0.21,0.49,0.74}
\usepackage[pagebackref,breaklinks,colorlinks,allcolors=cvprblue]{hyperref}

\def\paperID{*****} 
\def\confName{CVPR}
\def\confYear{2026}
\title{Gesture2Music: A Low-Latency Real-Time Framework for Continuous Gesture-Driven Music Generation}

\author{
  Rathinaraja Jeyaraj$^{1}$, Barathi Subramanian$^{1}$, Kapilya Gangadharan$^{2}$, Anand Paul$^{3}$ \\
  $^{1}$Stanford University, USA, $^{2}$Saveetha Institute of Medical and Technical Sciences, India \\ $^{3}$LSU Health Sciences Center New Orleans, USA \\
  {\tt\small \{rajaj,barathi1\}@stanford.edu, kapilyag@gmail.com, apaul4@lsuhsc.edu}
}

\begin{document}
\maketitle
\begin{abstract}
Gesture-driven music generation is an emerging human-computer interaction paradigm for touch-free and expressive musical interaction. However, many existing approaches treat the task as isolated gesture classification or map gestures to symbolic outputs such as MIDI followed by a separate rendering stage, which limits temporal continuity and real-time responsiveness. This work presents Gesture2Music, a low-latency streaming framework for continuous gesture-driven music generation from live webcam feed. The system processes sequences of body and hand landmarks and uses a causal temporal convolutional network (TCN) to predict note-level musical control events, including pitch, octave, onset, sustain, amplitude, and activity state. Because available gesture-note datasets typically contain only isolated single-note recordings rather than continuous performance sequences, a synthetic stream generation strategy is introduced to construct continuous gesture streams by concatenating single-note clips and deriving heuristic temporal event labels. Temporal consistency and spectral proxy losses are further used to reduce prediction jitter and encourage audio-consistent outputs. During inference, predicted musical events are rendered into continuous music using predefined note samples with rhythmic quantization and scale-constrained filtering for improved musical stability. Experiments on a custom gesture-to-music dataset with 21 gesture-note classes spanning seven tones across three pitch levels demonstrate stable real-time performance, low inference latency of 30\,ms, and improved temporal continuity.
\end{abstract}    
  \vspace{-25pt}
  \section{Introduction}
Gesture recognition is the task of estimating human body or hand motion patterns from visual observations and mapping them to pre-defined semantic actions or control signals \cite{ref1}. In human-computer interaction (HCI), gesture recognition has been widely studied as a non-contact interface that enables users to communicate with digital systems through natural movement rather than physical input devices. This paradigm has been explored across healthcare, education, and interactive systems \cite{ref2}–\cite{ref7}. Within music-related applications, gesture-based interfaces have been explored for tasks such as playback control \cite{ref8}, emotion-aware recommendation \cite{ref9}, and instrument-specific motion analysis \cite{ref10}. More recently, gesture-driven music generation has emerged as a promising direction for touch-free creative interaction, where users produce notes, rhythms, or short musical phrases through body and hand movements. Such systems are particularly attractive in settings where physical instruments are impractical or inaccessible, including immersive HCI, embodied learning, and rehabilitation-oriented creative environments. Despite this potential, reliable real-time gesture-to-music generation remains technically challenging. 

\begin{figure}[!t]  
	\centering
	\includegraphics[width=0.41\textwidth]{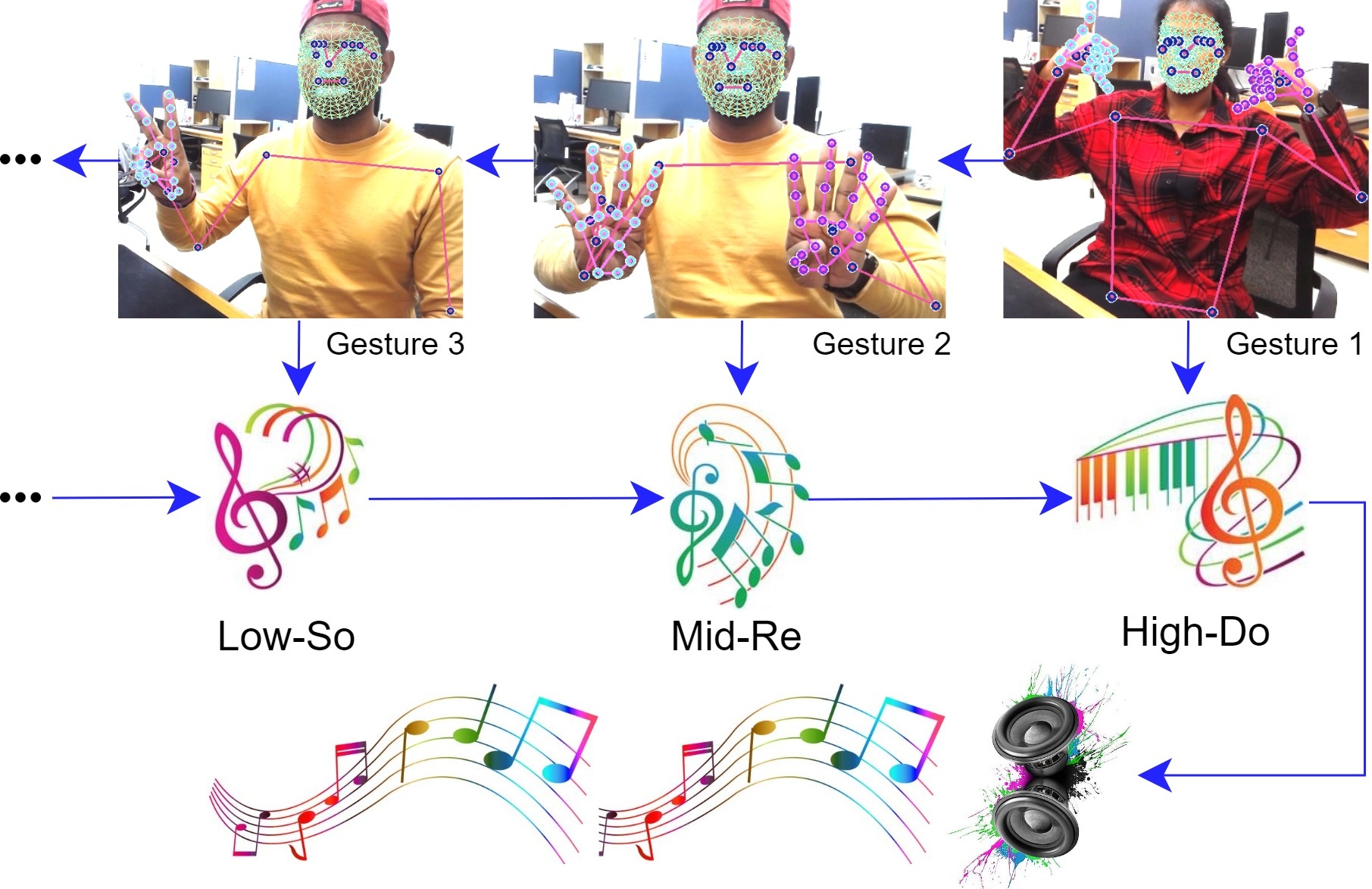}
	\caption{Vision-based gesture-recognition system for real-time music generation.}
	\label{fig:1}
    \vspace{-18pt}
\end{figure} 

A practical system must recognize gestures accurately and maintain temporal stability under variations in viewpoint, lighting, self-occlusion, and background clutter while producing low-latency audio responses that feel continuous and musically coherent. Many prior approaches treated gesture understanding as isolated classification, where each gesture is mapped independently to a label or command \cite{ref11}, \cite{ref12}. Other systems predicted symbolic outputs, such as MIDI events, and delegated acoustic rendering to a separate downstream stage \cite{ref13}. A second limitation is data availability. Gesture-to-note datasets are typically collected as isolated examples in which each gesture corresponds to a single musical target. These datasets are suitable for discrete recognition, but they do not provide continuous performance streams with temporal event annotations required for streaming gesture-to-audio generation. As a result, a model trained only on isolated gesture-note pairs may perform well in offline classification while failing to produce stable note onset, sustain, release, and transition behavior in real-time use. To address these limitations, this work presents Gesture2Music (Figure \ref{fig:1}), a low-latency streaming framework for continuous gesture-driven music generation from live video. 

The proposed system follows a six-stage pipeline. First, video frames were acquired from a standard camera and processed using MediaPipe \cite{ref14} to extract body and hand landmarks. Second, short rolling windows of landmark sequences were maintained in a first-in-first-out temporal buffer, enabling the model to operate on recent motion history rather than single frames. Third, the buffered sequence was processed by a causal temporal convolutional network (TCN) \cite{ref15} composed of stacked depthwise temporal convolution blocks with increasing dilation factors. The use of causal convolutions ensured strictly forward-looking prediction without future-frame access, which is essential for real-time deployment. Fourth, the learned temporal representation was passed to multi-task prediction heads that estimated note-level musical control events, including pitch, octave, onset, sustain, activity state, and amplitude. This event-based formulation separated musical state prediction from final audio playback and allowed the model to represent both discrete and continuous aspects of performance. Because the training data consisted of isolated gesture-note examples rather than naturally continuous musical streams, a synthetic stream generation strategy is introduced. This procedure concatenated single-note gesture clips into longer pseudo-continuous sequences and derived heuristic temporal supervision for event transitions. The training setup exposed the network to note boundaries, short-duration sustain patterns, and changing gesture contexts that more closely resembled real streaming interaction. To further improve temporal reliability, temporal consistency and spectral proxy loss were incorporated to reduce frame-to-frame jitter and encourage event predictions aligned with stable audio behavior.

During inference, raw event predictions are processed with three musical heuristics to improve perceptual stability before rendering the music. A confidence-aware pentatonic bias reduced the probability of dissonant note outputs under low-confidence conditions. A transition stabilization module blended the current prediction with a Markov-style transition prior derived from recent note history to suppress erratic switching. A BPM-based quantization \cite{ref16} stage synchronized note triggering through a beat-aligned queue, improving rhythmic consistency during streaming playback. Finally, the refined event sequence controlled a real-time sample-based rendering engine that retrieved note waveforms from a predefined audio bank, applied amplitude scaling and release decay, and streamed short audio chunks to the output device. The model runs at approximately 25-30 ms inference latency, while the full interactive pipeline operates at approximately 60-70 ms loop latency, supporting responsive real-time interaction. 

\begin{figure*}[!b]
  \vspace{-10pt}
	\centering
	\includegraphics[width=0.92\textwidth]{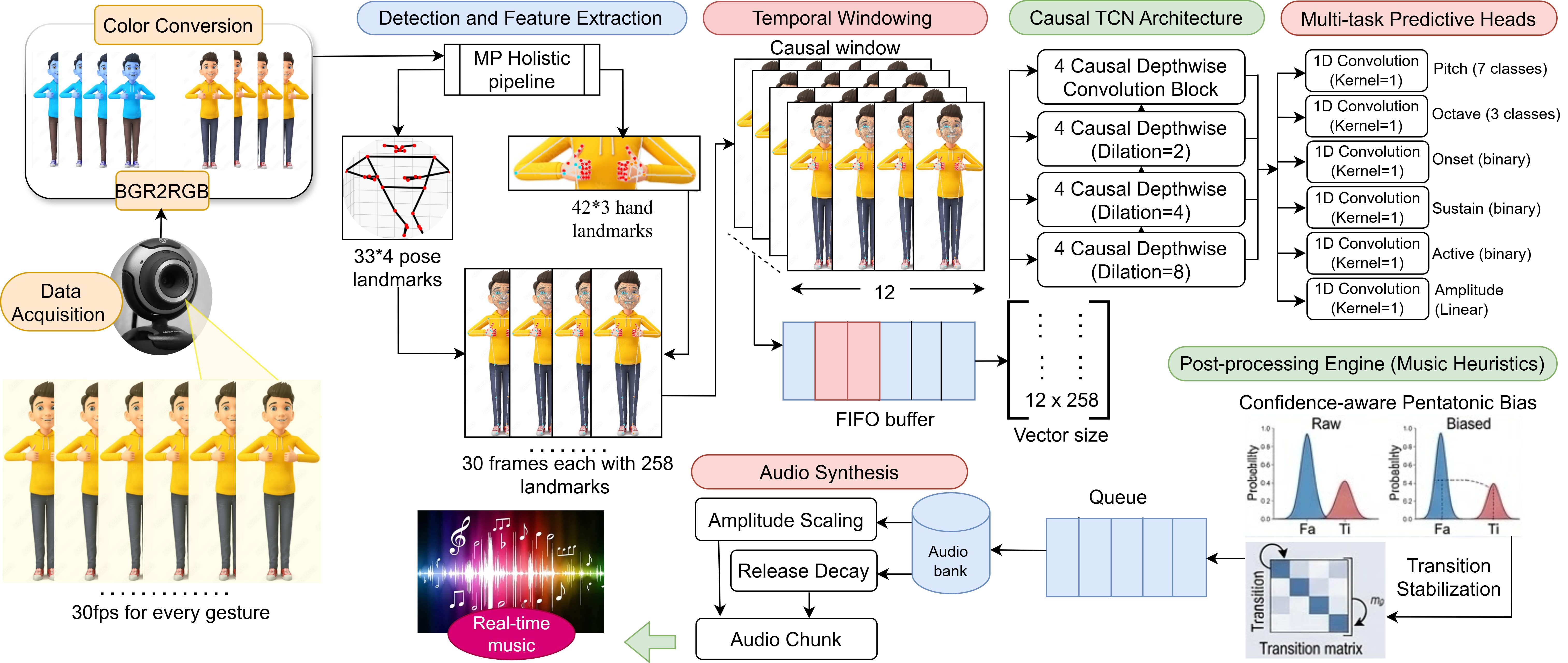}
	\caption{The proposed Gesture2Music framework.}
	\label{fig:2}
\end{figure*} 


To evaluate the framework, a custom dataset with five volunteers was collected. The dataset covered seven standard musical notes across three octave levels, yielding 21 gesture-note classes in total. Each class was represented by a unique gesture performed over 30 frames, with body and hand landmarks extracted per frame. Thirty samples were collected for each class per participant, resulting in 630 samples per volunteer, 3150 raw gesture clips overall, which were further converted into synthetic streaming sequences for model training. Experiments compare the proposed causal TCN against recurrent streaming baselines based on GRU and LSTM under the same training protocol. The results demonstrated that the proposed approach produced more stable real-time music generation, lower latency, and better temporal continuity than static gesture baselines. Unlike symbolic input modalities such as text or MIDI, visual gesture input provides continuous kinematic control signals including velocity, acceleration, articulation, and amplitude modulation. These fine-grained motion cues enable expressive real-time control that extends beyond discrete symbolic representations. The objective of this work is therefore not only note prediction but embodied interaction, where musical output emerges directly from human motion dynamics. Overall, we frame gesture-driven music generation as a streaming event prediction problem rather than a discrete classification task. The main contributions of this work are as follows: we 

\begin{itemize} [noitemsep,topsep=0pt]
    \item introduced a causal event-prediction model that maps landmark sequences to note-level musical control events.
    \item developed a synthetic stream construction and stabilization strategy to support temporally coherent real-time music generation from isolated gesture-note samples.
\end{itemize}  


The remainder of this paper is organized as follows. Section 2 reviews related work, Section 3 presents the proposed Gesture2Music framework, Section 4 reports the experimental results, and Section 5 concludes with limitations and future directions.

\section{Related work }  

\textit{Gesture recognition and real-time motion modeling:} Gesture recognition has been extensively studied in touch-free HCI, where body and hand movements are mapped to semantic actions or control signals \cite{ref17}-\cite{ref19}. Landmark-based representations have become useful for real-time systems as they provide compact geometric motion descriptors while avoiding the cost of full-frame visual modeling. Since gestures are inherently temporal, existing works employed recurrent networks and temporal convolutional models to capture motion dynamics \cite{ref20}. Real-time tracking frameworks such as MediaPipe further enabled efficient extraction of pose and hand landmarks for downstream sequence modeling \cite{ref21}.  \\[3 pt]
\textit{Gesture-driven music generation:} Gesture-driven music systems have long explored mappings from movement to sound events, synthesis parameters, or symbolic note representations \cite{ref11}, \cite{ref12}. Earlier approaches were often based on train-by-demonstration or supervised mappings, while more recent methods used sequence models to predict symbolic music from motion-derived features \cite{ref13}, \cite{ref8}, \cite{ref22}. Although they demonstrated the viability of gesture-based musical interaction, many treated gestures as isolated inputs and stitched the output to render final audio. This design often limited temporal continuity during live interaction.  \\[3 pt]
\textit{Low-latency musical stabilization and rendering:} Real-time gesture-to-music interaction requires accurate prediction; it requires stable note transitions, rhythmically coherent triggering, and low-latency audio output. Prior interactive music systems frequently relied on smoothing, transition priors, or beat-aware control to improve perceptual stability \cite{ref11}, \cite{ref12}. 


\section{Methodology}
This section presents the Gesture2Music framework for low-latency gesture-driven music generation. The system converts a live video stream into note-level musical control events and renders audio in real-time using a sample-based synthesis engine. The framework (Figure \ref{fig:2}) consists of six stages: (1) video acquisition and landmark extraction, (2) temporal windowing, (3) causal temporal modeling, (4) multi-task event prediction, (5) musical post-processing, and (6) real-time audio rendering. The design emphasizes causal inference, temporal stability, and low deployment latency.

\begin{figure*}[!b]
  \vspace{-10pt}
	\centering
	\includegraphics[width=1\textwidth]{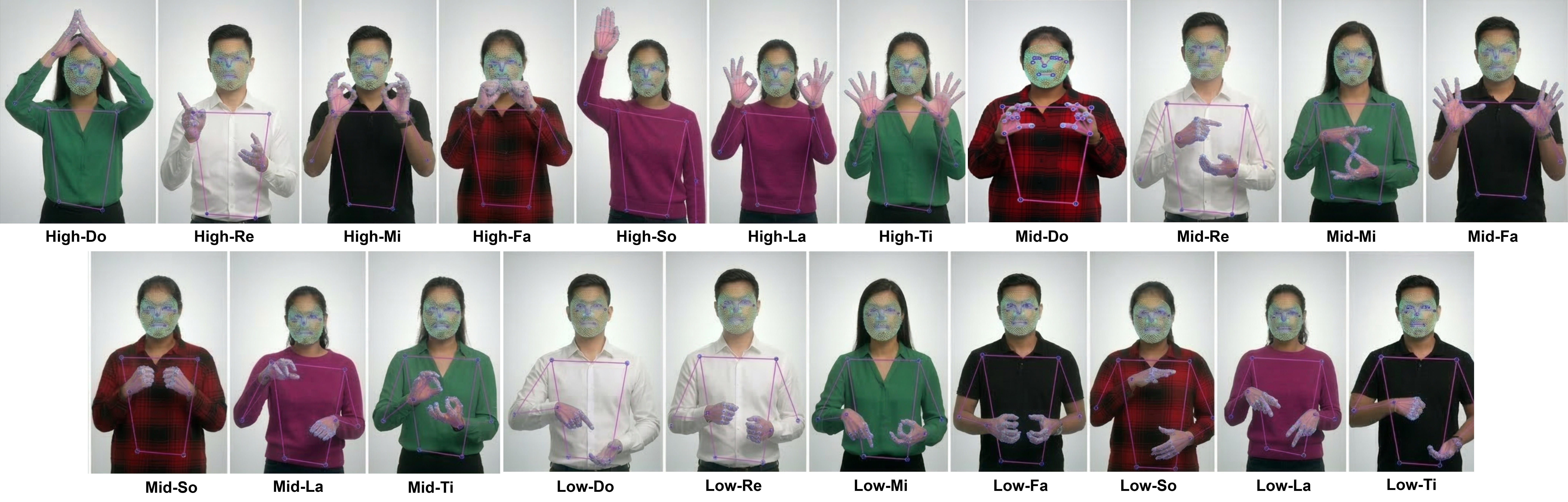}
	\caption{Sample poses for 21 different classes.}
	\label{fig:3}
\end{figure*} 

\subsection{Problem Formulation}

Let $I_t \in \mathbb{R}^{H \times W \times 3}$ denote the RGB frame captured at timestep $t$. Given a causal observation window of length $T$, $I_{t-T+1:t} = \{I_{t-T+1}, I_{t-T+2}, \dots, I_t\}$, the objective is to estimate a gesture-conditioned musical control state and generate a corresponding audio response in real-time. The musical control state is defined as

\begin{equation}
\mathcal{E}_t = \{\hat{\mathbf{p}}_t, \hat{\mathbf{o}}_t, \hat{u}_t, \hat{s}_t, \hat{c}_t, \hat{a}_t\},
\end{equation}
where, $\hat{\mathbf{p}}_t \in \mathbb{R}^{7}$ is the predicted pitch distribution over the notes $\{\text{Do, Re, Mi, Fa, So, La, Ti}\}$, $\hat{\mathbf{o}}_t \in \mathbb{R}^{3}$ denotes the octave distribution over $\{\text{Low, Mid, High}\}$, $\hat{u}_t$ is the onset probability, $\hat{s}_t$ is the sustain probability, $\hat{c}_t$ denotes the activity state probability, and $\hat{a}_t \in \mathbb{R}_+$ represents the predicted amplitude. The model therefore learns a causal mapping 
\begin{equation}
f_{\theta}: I_{t-T+1:t} \rightarrow \mathcal{E}_t,
\end{equation}
where $\theta$ represents the model parameters. Unlike conventional gesture-to-audio pipelines that perform isolated gesture classification followed by external audio playback, the proposed formulation predicts structured musical control events suitable for continuous interaction.
\vspace{-1.5pt}
\subsection{Data Acquisition and Labels}
Because no public dataset is available for gesture-driven music generation, a custom dataset was collected for this study. The dataset consisted of seven musical notes, $\mathcal{N}=\{\text{Do, Re, Mi, Fa, So, La, Ti}\}$, performed at three octave levels, $\mathcal{O}=\{\text{Low, Mid, High}\}$, resulting in $|\mathcal{N}| \times |\mathcal{O}| = 21$ gesture-note classes. Each class was associated with a unique gesture (Figure \ref{fig:3}) and recorded using a webcam at 30 frames/s from five volunteers under varying lighting conditions, viewpoints, and background settings. Each gesture instance was captured as a temporal sequence of 30 frames. During model training and inference, causal temporal windows of length $T=12$ were extracted using a sliding-window strategy. For each participant, 30 samples were recorded per class, yielding $21 \times 30 = 630$ samples per participant and $5 \times 630 = 3150$ samples in total. Inter-note pauses were represented by short neutral segments without an assigned note label and with inactive event targets. Reference audio recordings in \texttt{.wav} format were collected for each note class using a sampled digital piano sound bank to construct the audio dictionary used during the rendering stage. To simulate continuous musical interaction, synthetic streaming sequences were generated by concatenating isolated gesture clips with short neutral pauses between notes. Temporal labels for onset, sustain, activity state, and amplitude were derived using heuristic rules to each clip, with amplitude normalized to $(0,1)$ to represent relative gesture intensity. Specifically, onset was assigned to the initial attack portion of each clip, sustain to the central steady segment, and amplitude was generated using a normalized attack–sustain–release envelope scaled to $[0,1]$; short inserted pause segments were labeled with zero onset, zero sustain, zero amplitude, and inactive state. Dataset splits were defined at the clip level prior to synthetic stream construction, ensuring that isolated gesture clips used to generate validation sequences were not reused in training streams.

\vspace{-1.5pt}
\subsection{Geometric Landmark Stream}
MediaPipe \cite{ref14} is used to estimate pose and hand landmarks from the observed RGB frames with 33 pose landmarks, 21 left-hand landmarks, and 21 right-hand landmarks. This yielded 75 landmarks per frame. Landmark coordinates were encoded into a feature vector $\mathbf{l}_t \in \mathbb{R}^{258}$, (33$\times$4+21$\times$3+21$\times$3) constructed by concatenating pose $(x,y,z,v)$ attributes with $(x,y,z)$ coordinates for the left- and right-hand landmarks. A causal input window was therefore represented as $ \mathbf{L}_{t-T+1:t} = [\mathbf{l}_{t-T+1}, \dots, \mathbf{l}_t] \in \mathbb{R}^{T \times 258}$, where $T$ denotes the temporal window size used for streaming prediction. This representation provided geometric information describing body posture and hand configuration.

\subsection{Temporal Windowing}
To capture temporal context, the framework maintained a rolling memory buffer over the most recent landmark vectors, using a causal window of length $T=12$ frames. A first-in-first-out (FIFO) queue constructed the streaming input window 
$\mathbf{X}_t = \mathbf{L}_{t-T+1:t} \in \mathbb{R}^{T \times D}$, where $D=258$. At each time step, a new landmark vector was appended and the oldest vector was removed. This sliding strategy enabled causal sequence modeling without accessing future frames.

\subsection{Causal TCN Architecture}
The temporal sequence $\mathbf{X}_t$ by a causal TCN with dilated causal convolutions, following the general design principle popularized in WaveNet \cite{ref23}. The network consisted of stacked depthwise temporal convolution blocks with dilation factors $d \in \{1,2,4,8\}$. For block $k$, the feature update was
\vspace{-3pt}
\begin{equation} 
\mathbf{H}^{(k)} =
\phi(\text{Conv}_{\text{causal}, d_k}(\mathbf{H}^{(k-1)})),
\end{equation} 
where $\phi(\cdot)$ denotes nonlinear activation and normalization. Left-side zero padding ensured that each prediction depended only on past observations. The final latent representation was defined by $\mathbf{z}_t = \text{Pool}(\mathbf{H}^{(K)})$. 

\subsection{Multi-Task Predictive Modeling}
From the latent representation $\mathbf{z}_t$, the model predicted musical control events through task-specific heads. \\[2 pt]
\textbf{Pitch Prediction:} To identify the musical note associated with the observed gesture, the model predicts a categorical distribution over the seven pitch classes.
\vspace{-3pt}
\begin{equation}
\hat{\mathbf{p}}_t =
\text{softmax}(W_p \mathbf{z}_t + b_p).
\end{equation}
\noindent
\textbf{Octave Prediction:} To model pitch height variations within the musical scale, an additional head predicts the octave level corresponding to the gesture.
\vspace{-3pt}
\begin{equation}
\hat{\mathbf{o}}_t =
\text{softmax}(W_o \mathbf{z}_t + b_o).
\end{equation}
\textbf{Event Probabilities:} To capture temporal note dynamics, the model estimates onset, sustain, and activity probabilities that regulate note triggering and persistence.
\vspace{-3pt}
\begin{align}
\hat{u}_t &= \sigma(W_u \mathbf{z}_t + b_u), \\
\hat{s}_t &= \sigma(W_s \mathbf{z}_t + b_s), \\
\hat{c}_t &= \sigma(W_c \mathbf{z}_t + b_c).
\end{align}
\textbf{Amplitude Prediction:} To modulate the loudness of the rendered audio, a regression head predicts a continuous amplitude parameter conditioned on gesture dynamics.
\begin{equation}
\hat{a}_t = \sigma(W_a \mathbf{z}_t + b_a).
\end{equation}
The training objective combines classification, event prediction, regression, and temporal regularization losses:
\vspace{-3pt}
\begin{equation}
\begin{aligned}
\mathcal{L} = \, & \lambda_p \mathcal{L}_{pitch} + \lambda_o \mathcal{L}_{octave} + \lambda_{on} \mathcal{L}_{onset} + \lambda_s \mathcal{L}_{sustain} \\
& + \lambda_a \mathcal{L}_{amp} + \lambda_c \mathcal{L}_{active} + \lambda_t \mathcal{L}_{temp} + \lambda_{sp} \mathcal{L}_{spec}.
\end{aligned}
\end{equation}
A temporal consistency regularizer reduces prediction jitter by penalizing abrupt changes between sequence predictions:
\vspace{-3pt}
\begin{equation}
\mathcal{L}_{temp} =
\frac{1}{B(T-1)C}\sum_{b=1}^{B}\sum_{t=2}^{T}\sum_{c=1}^{C}
\left(\hat{y}_{b,t,c}-\hat{y}_{b,t-1,c}\right)^2,
\end{equation}
where $B$ denotes the batch size, $T$ the sequence length, and $C$ the number of output channels. In practice, this consistency regularizer is applied to both the pitch and octave prediction sequences. Although the prediction heads are optimized separately, they share the same latent representation produced by the causal TCN, which captures short-term temporal context from preceding gesture frames. Thus, temporal dependencies between musical events are modeled through the shared backbone. Additional temporal stability is encouraged by the sequence-level consistency loss during training and by the post-processing stage during inference. However, the current formulation does not explicitly model conditional dependencies between output variables. Loss weights were empirically selected using small validation sweeps to balance convergence stability across tasks.
\vspace{-1.5pt}
\subsection{Post-Processing Engine }
Predictions were refined via three stabilization modules. \\[3 pt]
\textit{Confidence-Aware Pentatonic Bias:} Low-confidence predictions were adjusted using a pentatonic prior to reduce unstable notes. \\[3 pt]
\textit{Transition Stabilization:} Predictions were blended with a Markov-style transition matrix
\begin{equation}
\bar{\mathbf{p}}_t =
(1-\eta)\hat{\mathbf{p}}_t + \eta \mathbf{T}_{n_{t-1}},
\end{equation}
where $\mathbf{T}$ denotes the transition matrix. The transition matrix acts as a first-order Markov prior encouraging musically consistent note transitions across adjacent time steps. \\[3 pt]
\textit{BPM Quantization:} A beat-synchronous queue released notes only at metronome boundaries (120 BPM) after stable prediction for $k$ frames.\\
\noindent
\begin{algorithm}[t!] 
\footnotesize
\caption{Gesture2Music streaming inference}
\label{alg:g2m_short}
\begin{algorithmic}
\Require Frame stream $\{I_t\}$, window length $T$, model $f_{\theta}$, audio bank $\mathcal{B}$
\Ensure Real-time audio output

\State Initialize FIFO buffer $\mathcal{M}$, previous note $n_{t-1}$, stability counter $c_{\mathrm{stable}}$, queue $\mathcal{Q}$

\For{each frame $I_t$}
    \State Extract landmark $\mathbf{l}_t \in \mathbb{R}^{258}$ using MediaPipe 
    \State Update FIFO buffer $\mathcal{M} \leftarrow \mathcal{M} \cup \mathbf{l}_t$
    \If{$|\mathcal{M}| < T$}
        \State \textbf{continue}
    \EndIf
    \State Form causal window $\mathbf{X}_t$
    \State Predict events: $ (\hat{\mathbf{p}}_t,\hat{\mathbf{o}}_t,\hat{u}_t,\hat{s}_t,\hat{c}_t,\hat{a}_t) \leftarrow f_{\theta}(\mathbf{X}_t)$
    
    \State Apply confidence-aware bias and transition stabilization to obtain $\bar{\mathbf{p}}_t$
    \State Decode note $n_t \leftarrow \arg\max \bar{\mathbf{p}}_t$
    \State Update stability counter for $n_t$
    \If{$c_{\mathrm{stable}} > k$}
        \State Queue current stable event
    \EndIf
    \If{beat boundary is reached and queue is not empty}
        \State Pop latest stable event
        \State Retrieve waveform $w_t \leftarrow \mathcal{B}(n_t)$
        \State Scale by amplitude and apply release decay 
        \State Stream audio chunk to output device
        \State Update previous note $n_{t-1} \leftarrow n_t$
    \EndIf
\EndFor
\end{algorithmic} 
\end{algorithm}
\vspace{-2pt}
\begin{figure*}[!t]
  \vspace{-10pt}
	\centering
	\includegraphics[width=0.96\textwidth]{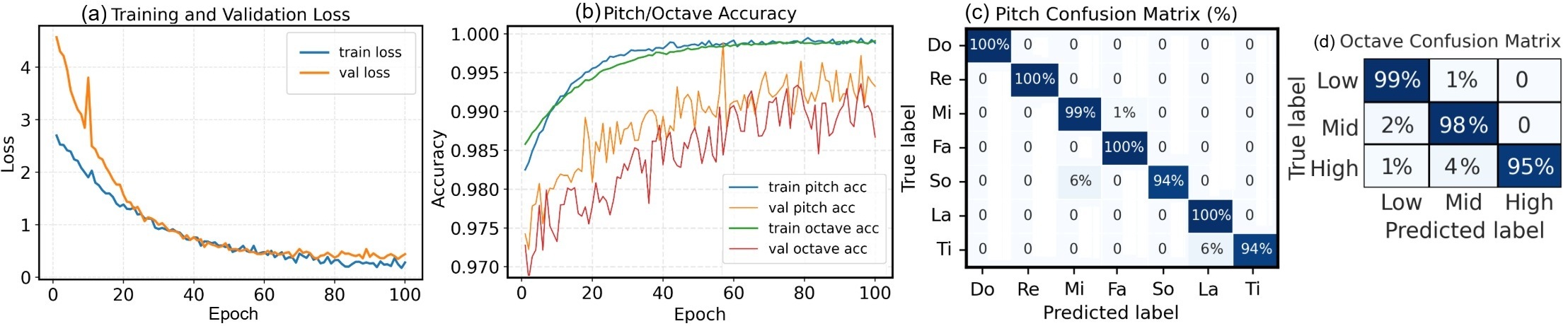}
	\caption{Training and validation loss curves, pitch and octave classification accuracy, and corresponding confusion matrices.}
	\label{fig:4}
      \vspace{-13pt}
\end{figure*} 

\vspace{-6pt}
\subsection{Sample-Based Audio Rendering}
The final stage converted note events into audio. Given the selected note $\bar{n}_t$ and amplitude $\hat{a}_t$, the renderer retrieved the waveform $w_t = \mathcal{B}(\bar{n}_t)$ from a predefined audio bank $\mathcal{B}$. The waveform was scaled as $\hat{s}_t = \hat{a}_t \cdot w_t$. Short audio chunks (800 samples) were streamed to the output device. A release-decay envelope was applied when notes became inactive to avoid abrupt cutoff artifacts. The neural model achieves approximately 25-30 ms inference latency, while the full system pipeline operates at approximately 60-70 ms loop latency.  Algorithm 1 summarizes the overall workflow. As the network predicts symbolic musical events rather than raw waveform samples, the rendering stage is instrument-agnostic. The same predicted note events can be mapped to different timbral sample libraries (e.g., piano, violin, flute) without modifying the learned gesture representation.

\section{Experiments and Results}
\subsection{Experimental Setup}
The proposed Gesture2Music framework was evaluated in a streaming gesture-to-music generation setting. The system mapped body and hand gestures from RGB video to note-level musical events, which were rendered into audio using a real-time sample-based synthesis engine. For each frame, pose and hand landmarks were extracted using MediaPipe and processed by a causal TCN over a fixed-length sliding temporal window ($T=12$), enabling short-term temporal modeling under strict real-time constraints. At each timestep, the model predicted six outputs: pitch over seven classes \{Do, Re, Mi, Fa, So, La, Ti\}, octave over three levels \{Low, Mid, High\}, onset, sustain, active state, and a continuous amplitude value. Pitch and octave were trained using cross-entropy loss, onset, sustain, and activity using binary cross-entropy, and amplitude using mean absolute error. As the dataset consisted of isolated gesture-note clips rather than continuous performances, synthetic gesture streams were constructed by concatenating clips with inserted pause intervals. Temporal supervision for onset, sustain, amplitude, and activity was generated using synthetic envelopes aligned with the concatenated streams. Validation streams were generated from a held-out subset of isolated gesture-note clips obtained by splitting the clip library before synthetic stream construction, so that the raw clips used in validation were not reused in training. During inference, predicted note events were converted into audio through a sample-based rendering that selects and modulates pre-recorded instrument samples based on predicted pitch, octave, and amplitude. Source code and demonstration videos are in the supplementary material. 

\begin{figure*}[!t]
  \vspace{-10pt}
	\centering
	\includegraphics[width=1\textwidth]{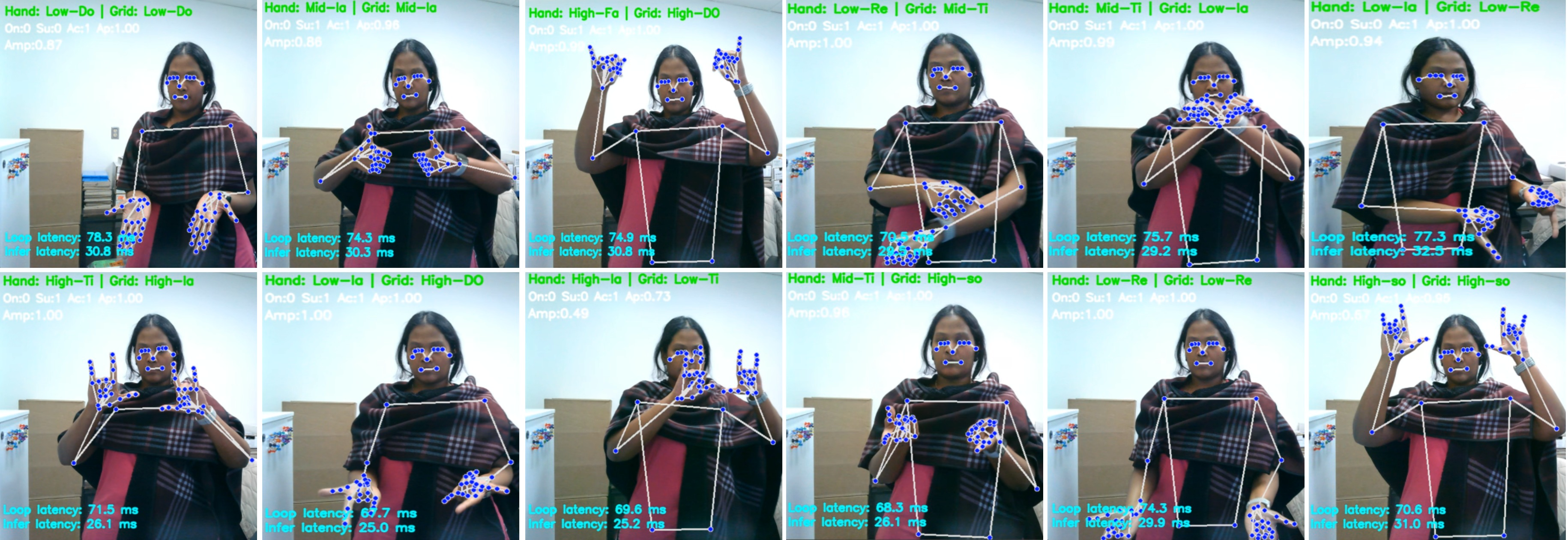}
	\caption{Sample real-time inference frames showing predicted hand notes and rendered grid notes.}
	\label{fig:5}
      \vspace{-10pt}
\end{figure*} 
 
\subsection{Training Behavior}
The training dynamics showed that the causal TCN learned the gesture-to-music mapping efficiently. Pitch prediction accuracy increased from approximately 97.5\% at the first epoch to nearly 99.9\% on the training set by the final epoch, as shown in Figure \ref{fig:4}, while validation pitch accuracy remained around 98\%--99.5\%. Octave prediction followed a similar trend, rising from about 98.2\% to 99.7\% on the training set, with validation accuracy remaining near 97\%--98\%. These results indicate that the model learned discriminative temporal features for both note identity and octave level within only a few epochs. The loss curves further showed stable optimization behavior. Training loss decreased steadily and approached near-zero values, while validation loss fluctuated between 0.6 and 1.5 but remained bounded. This gap likely reflected the variability introduced by synthetic stream generation, especially at gesture boundaries and pause transitions. Despite this, high validation accuracy indicates strong generalization for the main classification tasks.


\subsection{Pitch and Octave Classification Performance}
 Figure \ref{fig:4} (c,d) shows the corresponding confusion matrices. The pitch confusion matrix exhibits strong diagonal dominance across all seven note classes. Several classes are predicted perfectly or near-perfectly, including Do (100\%), Re (100\%), Fa (100\%), and La (100\%), while Mi reaches 99\%. The main errors are concentrated in two classes. So is predicted correctly in 94\% of cases, with its primary confusion occurring with Mi (6\%). Similarly, Ti is predicted correctly in 94\% of cases and is occasionally confused with La (6\%). These errors are concentrated between musically nearby classes and likely reflect similarities in gesture trajectories during transitional motion. The octave confusion matrix also shows strong performance across all three octave levels, with accuracies of 99\%, 98\%, and 95\% for the low, mid, and high octaves, respectively. Most predictions lie on the diagonal, and the remaining errors occur primarily between adjacent octave levels, particularly when high-octave gestures are predicted as mid octave (4\%) and when mid-octave gestures are predicted as low octave (2\%). This pattern is consistent with ambiguous vertical hand placement during rapid gesture transitions. Overall, the confusion matrices indicate that the temporal model reliably maps gesture dynamics to discrete musical attributes, with the remaining errors concentrated in a small number of similar pitch and octave classes. Sample qualitative outputs are shown in Figure \ref{fig:5}.

\begin{table}[b!]
  \vspace{-12pt}
\centering  
\scriptsize
\caption{Ablation study evaluating the impact of backbone architecture (TCN, GRU, LSTM), input representation (pose+hands vs. hands only), and temporal window size $T$ on validation pitch and octave prediction accuracy.}
\label{tab:1}
\resizebox{0.5\textwidth}{!}{%
\begin{tabular}{|l|c|c|c}
\hline
\multicolumn{1}{|c|}{\textbf{Model Variant}} & \textbf{Input}          & \textbf{Val. Pitch Acc (\%)} & \multicolumn{1}{c|}{\textbf{Val. Octave Acc (\%)}} \\ \hline
TCN (Pose+Hands)                             & Pose+Hands              & 97.90                        & \multicolumn{1}{c|}{97.89}                         \\ \hline
GRU (Pose+Hands)                             & Pose+Hands              & 94.26                        & \multicolumn{1}{c|}{95.68}                         \\ \hline
LSTM (Pose+Hands)                            & Pose+Hands              & 94.7                         & \multicolumn{1}{c|}{96.39}                         \\ \hline
TCN (Hands Only)                             & Hands                   & 96.39                        & \multicolumn{1}{c|}{96.56}                         \\ \hline
\textbf{Window size ($T$)}                   & \textbf{Pitch Acc (\%)} & \textbf{Octave Acc (\%)}     & \multicolumn{1}{l}{\multirow{4}{*}{}}              \\ \cline{1-3}
\multicolumn{1}{|c|}{8}                      & 97.1                    & 97                           & \multicolumn{1}{l}{}                               \\ \cline{1-3}
\multicolumn{1}{|c|}{12}                     & 97.9                    & 97.9                         & \multicolumn{1}{l}{}                               \\ \cline{1-3}
\multicolumn{1}{|c|}{16}                     & 97.6                    & 97.4                         & \multicolumn{1}{l}{}                               \\ \cline{1-3}
\end{tabular}
}
\end{table}

\subsection{Ablation Studies}
\vspace{-2pt}
To analyze the contribution of the major design choices in the proposed framework, we conducted ablation experiments focusing on the temporal backbone, input representation, and streaming window size. Unless otherwise specified, all experiments were trained under identical settings and evaluated using validation pitch and octave classification accuracy. \\[2 pt]
\textbf{Backbone comparison.}
We first evaluated the effect of the temporal modeling architecture by comparing the proposed Temporal Convolutional Network (TCN) with recurrent baselines based on GRU and LSTM using the same pose-and-hand landmark input representation and a fixed temporal window of $T=12$. As shown in Table \ref{tab:1}, the TCN model achieved the best performance with 97.90\% pitch accuracy and 97.89\% octave accuracy. In contrast, the GRU and LSTM models obtained lower pitch accuracies of 94.26\% and 94.70\%, respectively, with octave accuracies of 95.68\% and 96.39\%. These results indicate that causal temporal convolutions are more effective for modeling short-term gesture dynamics in streaming gesture-to-music prediction than recurrent architectures.\\[2 pt]
\textbf{Input representation.}
We further examined the importance of the landmark representation by comparing a hands-only input with the full pose-plus-hands representation using the TCN backbone. The hands-only configuration achieved 96.39\% pitch accuracy and 96.56\% octave accuracy, whereas the pose-plus-hands representation improved performance to 97.90\% and 97.89\%, respectively. This improvement suggests that upper-body pose provides additional contextual information beyond local hand geometry, which helps disambiguate gestures that differ in arm placement or global motion patterns.\\[2 pt]
\textbf{Temporal window size.}
Finally, we analyzed the influence of the temporal context window by varying the streaming window size over $T \in \{8,12,16\}$ while keeping the TCN backbone and pose-plus-hands input fixed. The best performance was obtained at $T=12$, achieving 97.9\% pitch accuracy and 97.9\% octave accuracy. Reducing the window to $T=8$ slightly decreased performance to 97.1\% and 97.0\%, indicating that shorter windows provide less temporal context for capturing gesture transitions. Increasing the window to $T=16$ produced 97.6\% pitch accuracy and 97.4\% octave accuracy, showing no clear improvement over $T=12$. Based on this trade-off between temporal context and streaming latency, the final system adopts a 12-frame causal window for real-time inference.

\begin{figure}[!b]
  \vspace{-10pt}
	\centering
	\includegraphics[width=0.43\textwidth]{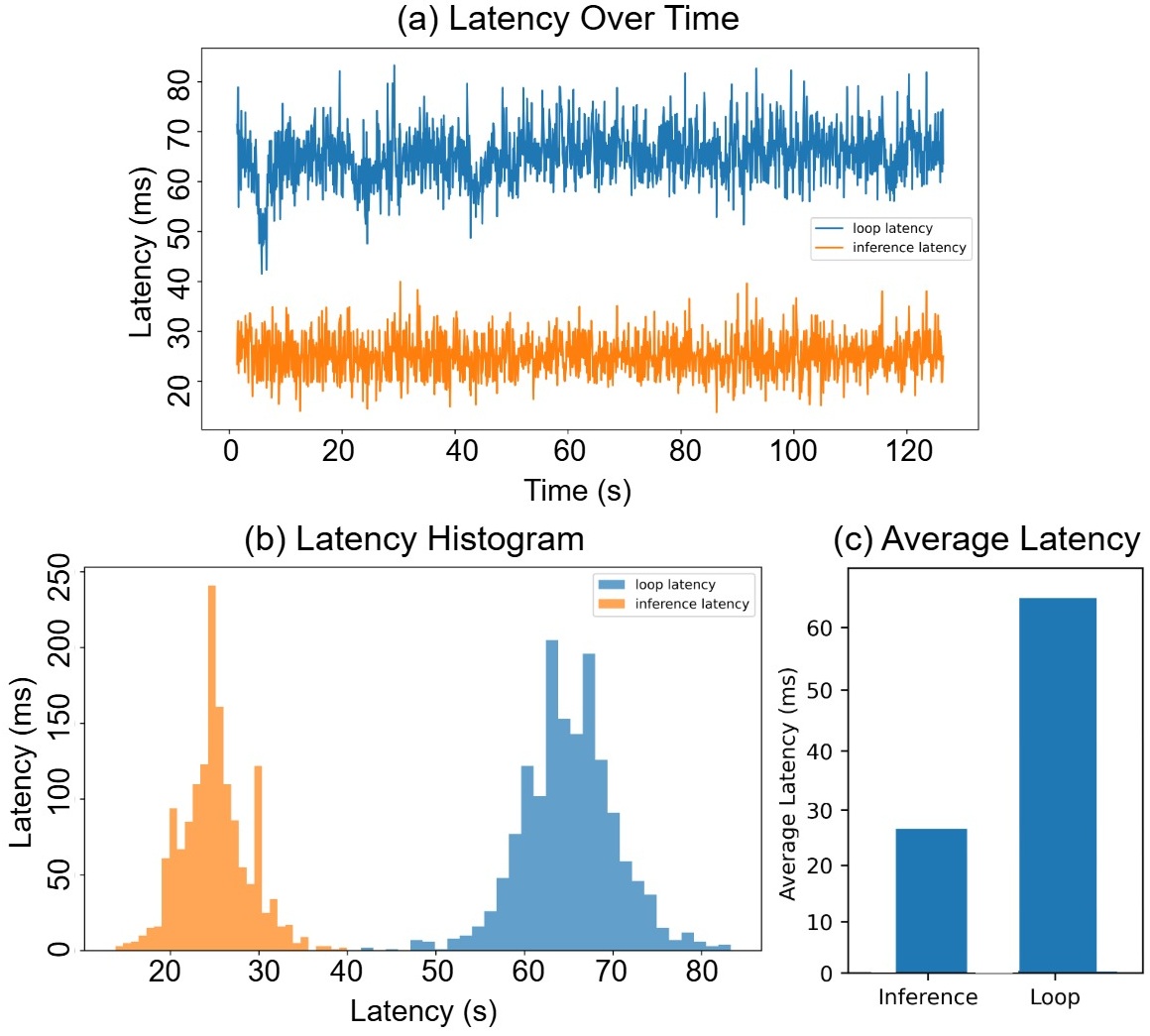}
	\caption{Real-time latency analysis.}
	\label{fig:6}
      \vspace{-15pt}
\end{figure} 

\subsection{Real-Time Latency Analysis}
Latency is a critical requirement for interactive gesture-driven music systems. The proposed system was therefore evaluated in terms of both neural network inference latency and total loop latency for the full processing pipeline, as shown in Figure \ref{fig:6}. Latency measurements were obtained using a single NVIDIA GPU with real-time webcam input pipeline and optimized streaming inference configuration. The average inference latency was approximately 25--30\,ms, while the average full loop latency was 60--70\,ms. The full loop measurement included landmark extraction, feature pre-processing, neural inference, post-processing, and audio rendering. These values remained well below the 100~ms threshold commonly regarded as acceptable for real-time human-computer interaction. The difference between inference latency and full loop latency further shows that the dominant runtime cost arose from the perception and rendering components rather than from the model itself. This result supports the practical viability of the proposed framework for interactive real-time use. Latency distribution and runtime stability were further analyzed using time-series traces, latency histograms, and average latency measurements (Figure \ref{fig:6}). Inference latency was concentrated between 25 and 30~ms, while full loop latency typically fell between 60 and 70~ms. However, both inference and loop latency remained stable throughout the evaluation interval, with only minor transient spikes. These results indicate that the streaming implementation maintained consistent timing behavior during continuous operation.

\begin{figure}[!b]
  \vspace{-10pt}
	\centering
	\includegraphics[width=0.48\textwidth]{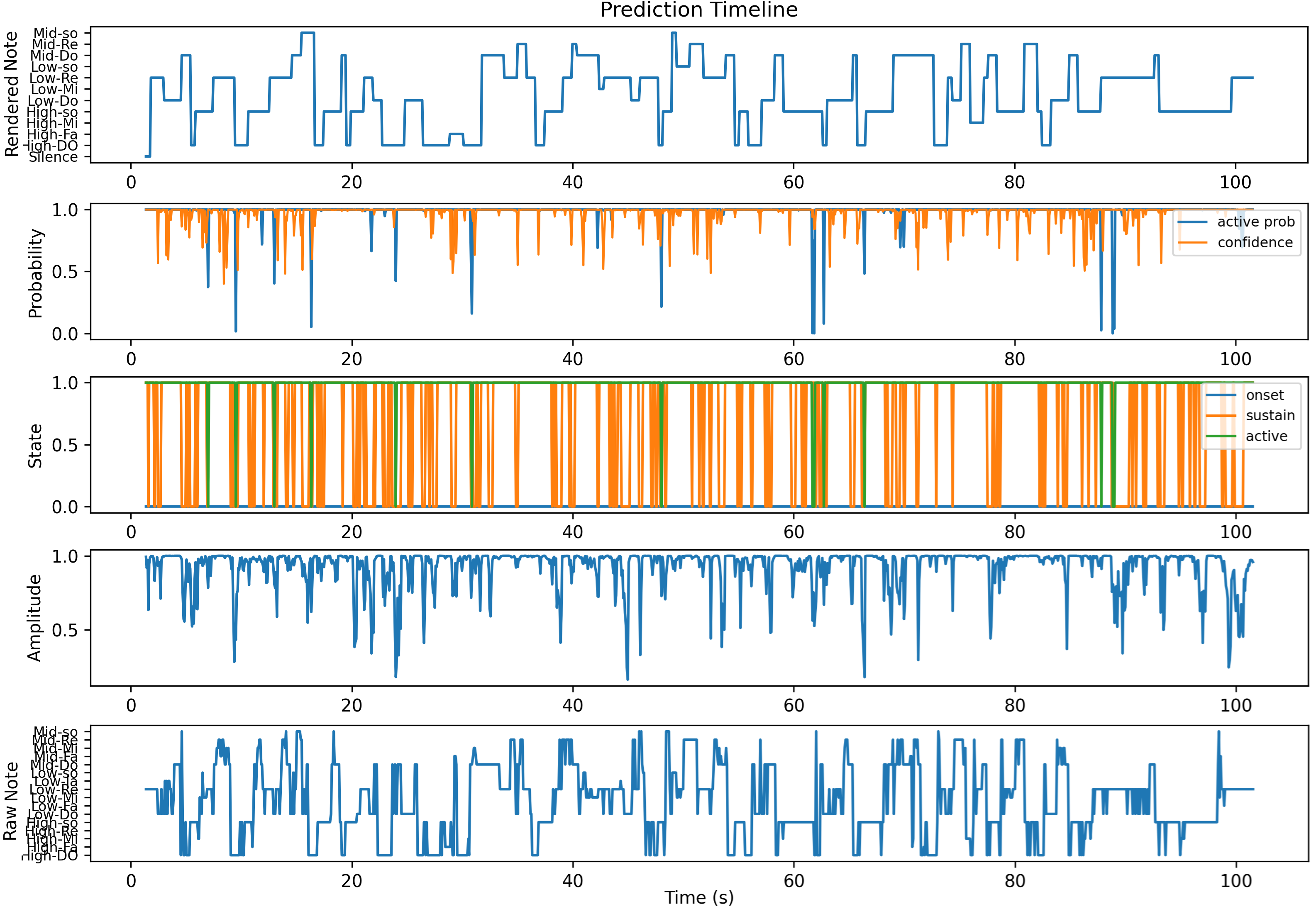} 
	\caption{Temporal prediction behavior of the proposed Gesture2Music system during real-time inference. The figure shows the rendered note sequence, model confidence and activity probabilities, event states (onset, sustain, active), amplitude prediction, and the raw predicted note sequence over time.}
	\label{fig:7}
      \vspace{-13pt}
\end{figure} 
 
\subsection{Temporal Prediction Behavior}
Temporal prediction behavior was examined using the timeline visualization of model outputs. Figure \ref{fig:7} shows a timeline visualization of predicted pitch classes, quantized output notes after stabilization, onset and sustain signals, activity state, and the resulting rendered note sequence. The predicted notes formed clear temporal segments corresponding to stable gesture intervals, while the rendered grid notes showed that the post-processing and quantization stages converted the raw predictions into musically coherent note transitions. The onset and sustain traces further indicated that the system captured temporal event structure rather than merely assigning static labels to each window. One limitation visible in the timeline is that the activity probability remains close to 1.0 for much of the sequence. However, the note predictions were stable over time and that the post-processing engine reduced erratic switching in the final output.


\subsection{Discussion}

Results demonstrate that the proposed system enables low-latency, real-time gesture-driven music generation. The model achieved high pitch and octave accuracy, with strong diagonal dominance in confusion matrices and rapid convergence during training. Runtime analysis showed 25–30 ms inference latency and 60–70 ms end-to-end loop latency, supporting responsive real-time interaction. Temporal visualizations indicated that predicted notes form stable segments that are converted into coherent rendered output through quantization and heuristic stabilization. However, the activity prediction head remained saturated across portions of the sequence, indicating that silence and note deactivation are modeled less effectively than pitch and octave. Improving temporal supervision around pauses and release transitions is therefore an important direction for future work. Although qualitative evaluation shows that gesture motion can influence pitch and timing, formal user studies evaluating learnability, controllability, and performer adaptation remain necessary to assess usability in HCI contexts.

The current dataset is constructed from isolated gesture-note clips concatenated to simulate continuous musical sequences. While this enables controlled supervision of onset, sustain, and amplitude signals, it does not fully capture expressive variability such as articulation, micro-timing, and performer-specific style. Future work will investigate continuous capture of unconstrained gesture sequences and larger performer populations to improve robustness across performance styles and recording conditions. Heuristic temporal labels provide scalable supervision but may not fully represent expressive musical timing.

\section{Conclusion}
This paper presents Gesture2Music, a low-latency streaming framework for continuous gesture-driven music generation from live video. Rather than treating the task as isolated gesture classification, the method formulates gesture-to-music interaction as structured event prediction. Using body and hand landmarks extracted from RGB frames, a causal temporal convolutional network predicts pitch, octave, onset, sustain, activity state, and amplitude under real-time causal constraints. To address the lack of continuous gesture-music datasets, synthetic streaming sequences are constructed from isolated gesture-note clips. A post-processing stage improves musical stability through transition smoothing and rhythmic quantization, enabling responsive and coherent touch-free musical interaction.




\end{document}